\overfullrule=0pt
\input harvmac

\lref\Bnon{
  N.~Berkovits,
  ``Pure Spinor Formalism as an N=2 Topological String,''
JHEP {\bf 0510}, 089 (2005),
[hep-th/0509120].
}

\lref\Bpure{
  N.~Berkovits,
  ``Super Poincare Covariant Quantization of the Superstring,''
JHEP {\bf 0004}, 018 (2000).
[hep-th/0001035].
}

\lref\explaining {
N.~Berkovits,
``Explaining the Pure Spinor Formalism for the Superstring,''
JHEP {\bf 0801}, 065 (2008).
[arXiv:0712.0324 [hep-th]].
}

\lref\DIS{
M. Dine, I. Ichinose, and N. Seiberg, ``$F$ Terms And $D$ Terms In String Theory,''  Nucl. Phys. {\bf B293} (1987) 253.
}

\lref\ADS{
J. J. Atick, L. J. Dixon, and A. Sen, ``String Calculation Of 
Fayet-Iliopoulos $D$-Terms In Arbitrary Supersymmetric Compactifications,''
Nucl. Phys. {\bf B292} (1987) 109-149.
}

 \lref\GrSe{
M. B. Green and N. Seiberg, ``Contact Interactions In Superstring Theory,'' Nucl. Phys. {\bf B299} (1988) 559.}

\lref\AS{
J. J. Atick and A. Sen, ``Two-Loop Dilaton Tadpole Induced By Fayet-Iliopouplos $D$ Terms In Compactified
Heterotic String Theories,'' Nucl. Phys. {\bf B296} (1988) 157-186.}

\lref\DD{
J. Distler and M. D. Doyle, ``World-Sheet Supersymmetry Without
Contact Terms,'' hep-th/9203058.}

\lref\DSW{
M. Dine, N. Seiberg, and E. Witten, ``Fayet-Iliopoulos Terms In String Theory,'' Nucl. Phys. {\bf B278} (1986) 769.
}

\lref\ChandiaIMA{
  O.~Chandia,
  ``The Non-minimal Heterotic Pure Spinor String in a Curved Background,''
[arXiv:1311.7012 [hep-th]].
}

\lref\BerkovitsAMA{
  N.~Berkovits and O.~Chandia,
  ``Simplified Pure Spinor b Ghost in a Curved Heterotic Superstring Background,''
[arXiv:1403.2429 [hep-th]].
}

\lref\osv{
  O.~Chandia,
  ``The b Ghost of the Pure Spinor Formalism is Nilpotent,''
Phys.\ Lett.\ B {\bf 695}, 312 (2011).
[arXiv:1008.1778 [hep-th]].
}

\lref\rennan{
  R.~L.~Jusinskas,
  ``Nilpotency of the b Ghost in the Non Minimal Pure Spinor Formalism,''
[arXiv:1303.3966 [hep-th]].
}

\lref\Wone{
  E.~Witten,
  ``More On Superstring Perturbation Theory,''
[arXiv:1304.2832 [hep-th]].
}

\lref\Wtwo{ 
E. Witten, ``The Feynman $i\varepsilon$ In String Theory,''
arXiv:1307.5124.
}

\lref\Previous{
A. Berera, ``Unitary String Amplitudes,'' 
Nucl. Phys. {\bf B411} (1994) 157-180.
}

\lref\DhPold{E. D'Hoker and D. Phong,  ``Lectures on Two-Loop Superstrings,'' Adv. Lect. Math. {\bf 1}  (2008) 85-123, hep-th/021111.}

\lref\DhP{E. D'Hoker and D. Phong,  ``Two-Loop Vacuum Energy For Calabi-Yau
Orbifold Models,'' Nucl. Phys. {\bf B877} (2013) 343-400.}

\def\bar{\overline}

\def\a{{\alpha}}

\def\Tr{{\rm{Tr}}}

\def\l{{\lambda}}

\def\lb{{\overline\lambda}}

\def\lb{{\overline\lambda}}
\def\b{{\beta}}

\def\g{{\gamma}}

\def\d{{\delta}}
\def\e{{\epsilon}}

\def\L{{\Lambda}}

\def\half{{1\over 2}}
\def\p{{\partial}}

\def\pb{{\overline\partial}}
\def\t{{\theta}}

\def\tb{{\bar\theta}}

\def\lb{{\bar{\lambda}}}

\Title{\vbox{\baselineskip12pt
\hbox{ }}}
{{\vbox{\centerline{Supersymmetry Breaking Effects}
\smallskip
\centerline{using the Pure Spinor Formalism of the Superstring}}} }
\bigskip\centerline{Nathan Berkovits\foot{e-mail: nberkovi@ift.unesp.br}}
\bigskip
\centerline{\it ICTP South American Institute for Fundamental Research}
\centerline{\it Instituto de F\'\i sica Te\'orica, UNESP - Univ. 
Estadual Paulista }
\centerline{\it Rua Dr. Bento T. Ferraz 271, 01140-070, S\~ao Paulo, SP, Brasil}
\bigskip
\centerline{and}
\bigskip\centerline{Edward Witten\foot{e-mail: witten@ias.edu}}
\bigskip
\centerline{\it School of Natural Sciences, Institute for Advanced Study}
\centerline{\it Princeton, NJ 08540, USA}
\vskip .3in

The SO(32) heterotic superstring on a Calabi-Yau manifold can spontaneously break supersymmetry
at one-loop order even when it is unbroken at tree-level.  It is known that calculating the supersymmetry-breaking
effects in this model gives a relatively accessible test case of the subtleties of superstring perturbation theory in the
RNS formalism.  In the present paper, we calculate the relevant amplitudes in the pure spinor approach to superstring perturbation
theory, and show
that the regulator used in computing  loop amplitudes in the pure spinor formalism leads to subtleties somewhat analogous
to the more familiar subtleties of the RNS approach.

\vskip .3in

\Date {April  2014}

\newsec{Introduction}

Like the Green-Schwarz (GS) formalism, the pure spinor 
formalism for the superstring \Bpure\ has the advantage over the 
Ramond-Neveu-Schwarz (RNS)
formalism that the worldsheet action is manifestly
spacetime supersymmetric. So there is no sum over spin structures in loop amplitudes
and Ramond-Ramond backgrounds can be conveniently described. However, in contrast to the GS
formalism, the worldsheet action in a flat background is quadratic in the pure spinor
formalism and $\kappa$-symmetry is replaced by a BRST symmetry, generated by a charge $Q$ whose square vanishes. 
So covariant quantization is straightforward and some
multiloop amplitudes can be explicitly computed. 

In a general loop computation, a potential subtlety arises.
In the pure spinor formalism, if one attempts to compute loop amplitudes using the manifestly
supersymmetric action that can be used to determine the spectrum, one has difficulty interpreting
the necessary integration over zero-modes.    To make sense of the zero-mode integrations,
one needs to add to the action a BRST-trivial regulator  term $\{Q,\dots\}$.  The regulator term is not spacetime supersymmetric
but since it is  BRST-trivial, its
supersymmetry variation is also BRST-trivial.  So formally one expects loop amplitudes
to be spacetime supersymmetric.  However, since a BRST-trivial term leads to an exact form
on moduli space, there is now an integration by parts on moduli space required to prove spacetime
supersymmetry.  There is  a possibility that the proof might fail in some cases because surface
terms will appear in the integration by parts.   When the proof does work, this might depend on a delicate
treatment of the region at infinity in moduli space.  

In the RNS approach to superstring perturbation theory, spacetime supersymmetry is not manifest and its proof
always involves an integration by parts on supermoduli space.  A delicate treatment is  needed and
requires working with super Riemann surfaces.  In many superstring compactifications with spacetime
supersymmetry, the subtleties do  not appear to arise in practical computations of low order.   A simple exception is the SO(32) heterotic string compactified
on a Calabi-Yau manifold.  Supersymmetry is unbroken at tree level, but generically it is spontaneously broken at the one-loop order
\DSW.
It turns out  that the supersymmetry-breaking effects that arise at 1- and 2-loop order\foot{In this model,
 superstring perturbation theory does not make sense to all orders: the model develops a dilaton tadpole
at 2-loop order and accordingly, generic
2-loop $S$-matrix elements will be infrared-divergent (except possibly after shifting the vacuum in a way that makes
string perturbation theory less straightforward).  Nonetheless, the low order amplitudes that do make sense in the model
give  a useful test case for superstring perturbation theory.} give interesting test cases for
superstring perturbation theory in the RNS formalism: they are simple enough that one can analyze them, but subtle enough
so that it is hard to do so correctly (and get the right answers) in an approach based on trying to reduce everything to ordinary rather
than super
Riemann surfaces.   Some of this was explained in the early literature \DIS\ADS\GrSe\AS\DD, and a more comprehensive analysis
has been made recently \Wone.

The purpose of the present paper is to analyze these same amplitudes in the pure spinor approach to superstring perturbation theory.
We will find that the regulator of the pure spinor approach leads to subtleties somewhat analogous to the subtleties of super Riemann
surfaces in the RNS formalism.  This result is not unexpected since otherwise it would not be possible for a model that is supersymmetric
at tree level to show spontaneous supersymmetry breaking at the 1-loop level.

In section 2 of this paper, we will review the pure spinor description of the heterotic superstring on a Calabi-Yau manifold with spin connection embedded in the gauge group. In a background with anomalous U(1) gauge field, the auxiliary scalar $D$ in the U(1) vector multiplet gets a one-loop expectation value $\langle D\rangle \neq 0$ which spontaneously breaks supersymmetry. 

In section 3, we compute in this background the one-loop mass splitting of bosons and fermions of charged multiplets. To regularize the one-loop two-point amplitude, one needs to put a cutoff so that the locations of the two vertex operators do not coincide.
And BRST invariance in the presence of this cutoff requires the inclusion of a contact
term. This contact term gives a contribution to the mass of charged bosons which is proportional to $\langle D\rangle$, but does not affect the mass of charged fermions.

In section 4, we compute in this background the two-loop cosmological constant (or vaccum energy). To regularize the zero-point two-loop amplitude, one needs to introduce a cutoff in the limit that the genus-two surface degenerates into two genus-one surfaces. BRST invariance again requires a contact term, and this contact term gives a contribution 
to the cosmological constant which is proportional to $\langle D\rangle^2$.

To some extent, our computations in this paper are experimental, since the subtleties of the pure spinor regulator have
not been fully explored.  Perhaps the examples we study will help in doing so.

\newsec{Review of Heterotic String}

\subsec{Worldsheet action}

Besides the spacetime $x^m$ variables for $m=0$ to 9, the left-moving worldsheet variables of the SO(32) heterotic string in 
the pure spinor formalism include the fermionic coordinates and conjugate momenta
$(\t^\a, p_\a)$ for $\a=1$ to 16, the pure spinor ghosts and their conjugate mometa $(\l^\a, w_\a)$, and the 
nonminimal bosonic and fermionic variables $(\bar\l_\a, \bar w^\a)$ and $(r_\a, s^\a)$ satisfying the constraints
\eqn\pure{\l\g^m\l = \lb\g^m\lb=0, \quad \lb\g^m r=0.}  (The nonminimal variables are not needed to construct a manifestly supersymmetric
model that can be quantized to give the superstring spectrum, but they facilitate the construction of the regulator.) 
The right-moving worldsheet variables
are the same as in the RNS formalism and include the fermionic spin-half variables $\xi^r$ for $r=1$ to 32 and the 
Virasoro ghosts $(\bar c, \bar b)$.

The pure spinor formalism can be defined in any consistent supergravity background and the worldsheet action is 
\eqn\curved{\eqalign{S ={1\over \a'} \int d^2 z &\biggl[ [\half \p Z^M (G_{MN}(Z)+ B_{MN}(Z)) + d_\a E^\a_N(Z) + {1\over 4} (\l \g_{ab} w)
\Omega^{ab}_N(Z)] \pb Z^N\cr
&+ [A_{Mi}(Z) \p Z^M + d_\a W^\a_i(Z)+ {1\over 4} (\l \g_{ab} w)  F^{ab}_i(Z)] \bar J^i +
\half \a' \Phi(Z) r \cr
&+ w_\a \pb\l^\a + \bar w^\a \pb\lb_\a +s^\a \pb r_\a +  \half\xi^r \p \xi^r +\bar b \p \bar c\biggr],\cr}}
where $Z^M = (x^m,\t^\a)$ are the $N=1$ $d=10$ superspace variables, $\half (\l \g_{ab}w)$ are the SO(9,1) Lorentz currents for the left-moving pure spinors and $a,b=0$ to 9 are
tangent-space vector indices, $\bar J^i= \xi^r T^i_{rs} \xi^s$ for $i=1$ to 496 are the SO(32) currents for the right-moving fermions, $T^i_{rs}$ are the SO(32) adjoint matrices, $(G_{MN},B_{MN}, E^\a_M, \Omega^{ab}_M, \Phi)$ are supergravity superfields
for the metric, antisymmetric tensor, spinor vierbein, spin connection and dilaton, 
$(A_{Mi}, W^\a_i, F^{ab}_i)$ are the super-Yang-Mills superfields for the gauge field, spinor field strength and vector field strength, and $r$ is the worldsheet curvature scalar. Although it will
be unnecessary in this paper, the nonminimal variables in $ \bar w^\a \pb\lb_\a +s^\a \pb r_\a$
can be made covariant with respect to local Lorentz transformations by coupling them to
the spin connection in a BRST-invariant manner as described in \ChandiaIMA\BerkovitsAMA.

Physical states are defined by dimension $(0,1)$ conformal primary fields of ghost-number 
$(1,0)$ which are in the cohomology of the left-moving BRST operator
\eqn\leftB{Q = \int dz ~(\l^\a d_\a + \bar w^\a r_\a)}
where the second term in \leftB\ ensures that the cohomology is independent of the nonminimal
variables. 
The BRST transformations of the superspace variables $Z^M = (x^m,\t^\a)$ are
$Q Z^M = \l^\a E_\a^M(Z)$ where $E_\a^M(Z)$ is the spinor component of the inverse supervierbein, so the BRST transformation of a scalar spacetime superfield
$\Phi(Z)$ is $Q\Phi(Z) = \l^\a D_\a\Phi(Z)$ where $D_\a = E_\a^M {{\p}\over{\p Z^M}}$ are the spacetime 
supersymmetric derivatives in the curved background.

For compactification on a Calabi-Yau manifold, the ten $x^m$ variables split into 
$(x^Q,y^J, \bar y^{\bar J})$ for $Q=0$ to 3 and $J,\bar J=1$ to 3, and the 32 right-moving $\xi^r$
variables split into $(\xi^J, \xi^{\bar J}, \xi^T)$ for $T=1$ to 26. It will be convenient to keep the 16-component SO(9,1) spinor index $\a$ for the spacetime spinors. The worldsheet
action in this background is
\eqn\curvedhet{S ={1\over \a'} \int d^2 z [\half \p x^Q \pb x_Q +  g_{J\bar J}\p y^J \pb y^{\bar J} + p_\a 
(\bar\nabla \t)^\a + w_\a (\bar\nabla\l)^\a + \xi^J (\nabla \xi)^{\bar J}}
$$+\half \xi^T\p \xi^T + \bar w^\a \pb\lb_\a  +s^\a \pb r_\a +\bar b \p \bar c + \dots],$$
where 
\eqn\none{\bar\nabla = \pb +{1\over 4}( \pb y^J \Omega_J^{ab}(y)\g_{ab}+\pb  y^{\bar J}\Omega_{\bar J}^{ab}(y) \gamma_{ab}),}
\eqn\ntwo{ \nabla = \p + \p y^J A_{Ji}(y)T^i + \p y^{\bar J} A_{\bar J i}(y)T^i - \half (p\g^{J \bar K}\t + w \g^{J\bar K} \l) F_{J\bar K i} T^i,}
$F_{J\bar K i}$ are auxiliary fields which will be relevant for supersymmetry breaking,
and $\dots$ denotes terms which are higher-order in $\t^\a$. The higher-order terms in $\t^\a$ will not be needed here and can be determined by requiring BRST invariance of the worldsheet action, or alternatively by solving the superspace equations of motion for the supergravity and super-Yang-Mills superfields in the Calabi-Yau background. 

In this background, $Q = \int dz ~ (\l^\a p_\a -\half (\l\g_Q\t)\p x^Q -\half (\l\g_a \t) (e^a_J(y) \p y^J +
e^a_{\bar J}(y) \p y^{\bar J})  + \bar w^\a r_\a + \dots )$
where $e^a_m$ is the vierbein and $\dots$ involves terms higher-order in $\t^\a$. And the
BRST transformations of the worldsheet variables are 
\eqn\brstcy{ Q x^Q =\half \l\g^Q\t, \quad Q y^J =\half (\l\g^a\t) e_a^J + \dots, \quad Q y^{\bar J}
=\half(\l\g^a\t) e_a^{\bar J} + \dots,\quad Q \t^\a = \l^\a + \dots}
where $\dots$ denotes terms higher-order in $\t^\a$.

\subsec{Vertex operators}

In the $d=10$ heterotic string, the supergravity and super-Yang-Mills onshell states
are described by the BRST-invariant vertex operators of dimension (0,1) and ghost-number (1,0): 
\eqn\vertex{V_{sugra} = \pb x^m \l^\a A_{\a m}(x,\t), \quad
V_{sYM} = \bar J^i \l^\a A_{\a i}(x,\t)}
where 
\eqn\expan{A_{\a m}(x,\t) = \half (\t\g^n)_\a (g_{nm}(x) + b_{nm}(x) +{2\over 3} (\t\g_n\rho_m(x)) +  \dots) ,}
\eqn\exg{A_{\a i}(x,\t) = \half(\t\g^m)_\a (a_{m i}(x) + {2\over 3} (\t\g_m \zeta_i(x)) -{1\over{16}}
 (\t\g_{mnp}\t) F^{np}_i(x) + \dots ),}
$(g_{mn}(x), b_{mn}(x), \rho_m^\a(x))$ and $(a_{mi}(x), \zeta^\a_i(x))$ are supergravity and 
super-Yang-Mills fields, and $\dots$ are terms higher-order in $\t^\a$ which are related by BRST invariance to the lower-order terms. For example, in a flat background, the BRST
transformation of the term
$(\t\g^m)_\a (\t\g_{mnp}\t) F^{np}_i(x)$ in \exg\ is cancelled by the BRST transformation
of the term $(\t\g^m)_\a a_{mi}(x)$
if and only if $F^{np}_i = \p^n a^p_i - \p^p a^n_i$. So $F^{mn}_i$ should be interpreted
as an auxliary field which is related at linearized level to the gauge field $a_{mi}$ by the
linearized equation of motion  $QV=0$.

In this paper, a central role will be played by the auxiliary field $D$ coming from dimensional
reduction of the $d=10$ Yang-Mills field strength $F^{mn}_i$ contracted with the Calabi-Yau metric as 
\eqn\Daux{D=g_{J \bar J} F^{J \bar J}_i T^i_{K \bar K} g^{K\bar K}.} 
The integrated vertex operator 
\eqn\integra{\int d^2 z ( g_{K\bar K}\xi^K \xi^{\bar K} )g_{J\bar J}(p\gamma^{J\bar J}\t + w\gamma^{J \bar J} \l)}
for this auxiliary field appears through the covariant
derivative \ntwo\ in the heterotic
sigma model action of \curvedhet. And the unintegrated
vertex operator for this auxiliary field in the pure spinor formalism can be obtained from the
term proportional to \Daux\ in \exg\ which is
\eqn\vd{V_D =  (g_{J\bar J}\xi^J \xi^{\bar J}) g_{K\bar K} (\l\g_m\t) (\t\g^{mK\bar K}\t)+ \dots,}
where $\dots$ denotes terms higher-order in $\t^\a$. This can be compared with the vertex operator for $D$ in
the RNS formalism, which is \DIS
\eqn\vdrns{V_D = c (g_{J\bar J}\xi^J \xi^{\bar J}) g_{K\bar K}\psi^K\psi^{\bar K}.}

To show that $V_D$ of \vd\ correctly describes the auxiliary field, note that the BRST transformations
of \brstcy\ imply
\eqn\qvd{Q V_D = 4 (g_{J\bar J}\xi^J \xi^{\bar J}) g_{K\bar K} (\l\g^K\t) (\l\g^{\bar K}\t)+ \dots .}
Defining $\Phi = \dots + \bar c V_D D(x)$ to be the string field and computing the quadratic term in the tree-level string field theory action ${1\over {2 g_s^2}}\langle \Phi Q \Phi \rangle$ using the measure
factor 
\eqn\measure{\langle \bar c \pb^2 \bar c~ (\l\g^m\t)(\l\g^n\t)(\l\g^p\t)(\t\g_{mnp}\t)\rangle =1,} 
one obtains the appropriate tree-level term in the field theory action
\eqn\treeD{S ={1\over{2 g_s^2}} \int d^4 x D(x) D(x).}

The action of \treeD\ can also be obtained by starting with the string field theory action
in a flat d=10 background and then setting $D = g^{J \bar J} F_{J\bar J}$ and restricting
the dependence on the six Calabi-Yau directions.
In a flat background, the quadratic term in the string field theory action is
\eqn\flatkin{S ={1\over{2 g_s^2}} \int d^{10} x (\half F_{mn} F^{mn} - \p_{[m} A_{n]} F^{mn}).}
which gives the usual d=10 Maxwell action after solving the equation of motion for the auxiliary
field $F_{mn}$ and plugging back into the action. 

At one loop, it will be shown in subsection 3.3 that the auxiliary
field $D$ couples linearly to the Calabi-Yau background field-strengths $F^{J\bar K}$ in the form
\eqn\ploop{S = p\int d^4 x D(x)}
where
\eqn\pinst{ p ={1\over{48\pi^3}} \int d^6 y \, \Tr_{SU(3)}(F\wedge F \wedge F).}
Combining the tree-level term of \treeD\ with the one-loop term of \ploop, the value of $D$ is shifted from the classical answer
$g^{J\bar J}F_{J\bar J}$ (which vanishes in Calabi-Yau compactification at the classical level) and one learns that
$D$ has the expectation value
\eqn\expv{\langle D\rangle = g_s^2 p}
which spontaneously breaks
spacetime supersymmetry when $p$ of \pinst\ is nonzero. In the next two sections, we will show how to compute the
effects of this supersymmetry breaking using the pure spinor formalism.

\newsec{One-Loop Mass Splitting}

In this section, the two-point one-loop amplitude of charged bosons will be shown to be nonzero after including a contact term required for BRST invariance. The analogous two-point one-loop amplitude of charged fermions related by supersymmetry is zero, implying that spacetime supersymmetry has been broken. 

\subsec{BRST-invariant prescription}

The two-point one-loop amplitude is naively computed in the pure spinor formalism by
\eqn\twoa{A =
 \int d^2 \tau \langle \bar c V^{(1)}(z_1) \int d^2 z_2 U^{(2)}(z_2) ~{\cal N}(y) |(\int \mu b)|^2 \rangle}
where $V^{(1)}$ is the unintegrated dimension (0,1) vertex operator of the first external state,
$U^{(2)}$ is the integrated dimension (1,1) vertex operator of the second external state, $(\int \mu b)$ is the insertion of the composite $b$ ghost contracted with the Beltrami differential $\mu$
dual to the Teichmuller parameter $\tau$,
and ${\cal N}$ is a regulator defined as ${\cal N} = e^{Q\Lambda}$ for some $\L$. The regulator ${\cal N}=e^{Q\Lambda}$ is needed to perform the functional integral over the noncompact
zero modes of the pure spinors
$(\l^\a, \lb_\a)$ and their conjugate momenta $(w_\a, \bar w^\a)$, and the precise form of 
$\L$ is irrelevant as long as the functional
integral is regularized and BRST-trivial states decouple. For example, choosing 
$\Lambda = -\rho (\lb_\a \t^\a + w_\a s^\a)$ where
$\rho$ is a positive constant implies that
\eqn\call{{\cal N} = e^{-\rho (\l^\a\lb_\a +\t^\a r_\a
+ w_\a \bar w^\a + d_\a s^\a)},}
which regularizes the functional integral over the
$(\l^\a,
\lb_\a, w_\a, \bar w^\a)$ zero modes for any choice of
$\rho$.

Because $U^{(2)}(z_2)$ may have a singularity when $z_2\to z_1$, one needs to introduce a cutoff $\e$ in the integral over $\int_\e d^2 z_2$ so that $z_2$ is integrated over the entire genus-one surface except for a small disk of radius $\e$ around $z_1$. Similarly, since the loop amplitude may be singular when the torus becomes infinitely thin, one needs to put a cutoff $\e'$ in the integration over the Teichmuller parameter. In the presence of these
cutoffs, BRST invariance is not manifest since $QU^{(2)} = \p V^{(2)}$ and $Qb = T$. So changing $\L$
in ${\cal N}$ will produce the term
\eqn\twoda{\d A =
 \int_{\e'} d^2 \tau \int_\e d^2 z_2  \langle \bar c V^{(1)}(z_1) U^{(2)}(z_2) ~(Q(\d\L)){\cal N}(y) |(\int \mu b)|^2 \rangle}
$$ =
- \int_{\e'} d^2 \tau  \int_\e d^2 z_2 \langle \bar c V^{(1)}(z_1)QU^{(2)}(z_2) ~(\d\L){\cal N}(y) |(\int \mu b)|^2 \rangle
$$
$$ + 
\int_{\e'} d^2 \tau\int_\e d^2 z_2 \langle \bar c V^{(1)}(z_1)  U^{(2)}(z_2) ~(\d\L){\cal N}(y) 
(\int\mu T)(\int \bar\mu \bar b)
 \rangle
$$
\eqn\twodb{=
- \int_{\e'} d^2 \tau \int_\e d^2 z_2 {\p\over{\p z_2}}\langle \bar c V^{(1)}(z_1)  V^{(2)}(z_2) ~(\d\L){\cal N}(y) |(\int \mu b)|^2 \rangle }
$$+
\int_{\e'} d^2\tau ~{\p\over{\p\tau}}  \int_\e d^2 z_2 \langle \bar c V^{(1)}(z_1) U^{(2)}(z_2) ~(\d\L){\cal N}(y) 
(\int \bar\mu \bar b)\rangle.$$

The second term in \twodb\ gives a surface term at the boundary of moduli space where the torus degenerates into an infinitely thin cylinder.
From a field theory point of view, one would say that the momentum flowing through this channel is an integration variable and thus is generically off-shell.
In general, in string theory, there is never a surface term arising from a degeneration  at which the momentum is generically off-shell.
The clearest explanation of this involves incorporating the Feynman $i\varepsilon$ in string theory (see \Wtwo\ on this point, 
and  see \Previous\ for an earlier discussion of the
Feynman $i\varepsilon$ in string theory).

The first term in \twodb\ gives 
\eqn\firstt{- \int_{\e'} d^2 \tau \langle \bar c V^{(1)}(z_1) \int_C d\bar z_2 V^{(2)}(z_2) ~(\d\L){\cal N}(y) |(\int \mu b)|^2 \rangle }
$$=  \int d^2 \tau \langle \bar c \Omega(z_1)  ~(\d\L){\cal N}(y) |(\int \mu b)|^2 \rangle$$
where $C$ is a contour of radius $\e$ around $z_2=z_1$ and
\eqn\defOm{\Omega(z_1) = - V^{(1)}(z_1) \int_C d\bar z_2 V^{(2)}(z_2).}

Since $V^{(1)}$ and $V^{(2)}$ are BRST-closed, $Q\Omega=0$. So $\Omega = Q\Sigma + V^{(3)}$ where 
$V^{(3)}$ is
in the BRST cohomology at ghost-number $(2,0)$. 
In the heterotic superstring, the BRST cohomology at ghost-number $(1,0)$ describes
physical vertex operators such as \vertex, and the BRST cohomology at ghost-number
$(2,0)$ describes the duals to these physical vertex operators.
If one assumes that $\d\L$ is a spacetime
scalar, $V^{(3)}$ must also be a spacetime scalar in order to contribute to \firstt. But
the only spacetime scalar in the BRST cohomology at ghost-number $(2,0)$ is the dual vertex
operator for the
dilaton, i.e.
\eqn\dil{V^{(3)} = (\t\g_{mnp}\t)(\l\g^n\t)(\l\g^p\t) \bar\p x^m.}
Note that the dual vertex operators in the super-Yang-Mills multiplet are either spacetime spinors or vectors. Since we are assuming the external
states are super-Yang-Mills states which do not carry $\bar\p x^m$ dependence, the vertex operator of \dil\ cannot be present in $\Omega$. So we will
have  $\Omega = Q\Sigma$
for some $\Sigma$ (in our examples, we will see explicitly what $\Sigma$ is) and
\eqn\twodd{\d A =  \int_{\e'} d^2 \tau \langle \bar c Q\Sigma(z_1) ~(\d\L){\cal N}(y) |(\int \mu b)|^2 \rangle }
$$=\int_{\e'} d^2 \tau \langle \bar c \Sigma(z_1) ~(Q\d\L){\cal N}(y) |(\int \mu b)|^2 \rangle $$
where the contribution from $Qb$ in \twodd\ can be ignored for the same reason that the second term in 
\twodb\ vanishes.

So although the amplitude prescription of \twoa\ is not manifestly BRST-invariant, it can
be modified to the BRST-invariant amplitude prescription 
\eqn\twoa{A =
 \int d^2 \tau \langle \bar c (V^{(1)}(z_1) \int_\e d^2 z_2 U^{(2)}(z_2) - \Sigma(z_1)) ~{\cal N}(y) |(\int \mu b)|^2 \rangle}
where $\Sigma$ is defined by 
\eqn\Sigdef{Q\Sigma = - V^{(1)}(z_1) \int_C d\bar z_2 V^{(2)}(z_2).}

\subsec{Mass splitting}

For charged bosons transforming as $26^{\pm 1}$ of $SO(26)\times U(1)$ which are associated with the (1,1)-form $\omega_{J\bar J}(Y)$, the unintegrated
vertex operators are
\eqn\verc{V_T^{(1)} = e^{ik\cdot  x} \xi_T \xi^J \omega_{J\bar J}(Y) (\l\g^{\bar J}\t) + \dots,\quad
V_{T'}^{(2)} = e^{-ik\cdot x} \xi_{T'} \xi^{\bar J} \omega_{\bar J J}(Y) (\l\g^{J}\t) + \dots,}
where $\dots$ denotes higher-order terms in $\t^\a$. Using the OPE
$\xi_T(\bar z_1) \xi_{T'}(\bar z_2) \to (\bar z_1 - \bar z_2)^{-1} \d_{T T'}$, one finds that
\eqn\Om{\Omega_{T T'} =\d_{T T'} (g_{J \bar J} \xi^J\xi^{\bar J}) g_{K\bar K}(\l\g^K\t)(\l\g^{\bar K}\t) + \dots}
 $$= 
Q ({1\over 4}\d_{T T'}(g_{J \bar J} \xi^J\xi^{\bar J}) g_{K\bar K}(\l\g_m\t)(\t\g^{m K\bar K}\t)  + \dots).$$
So 
\eqn\sigb{\Sigma_{T T'} ={1\over 4}\d_{T T'} V_D + \dots}
satisfies \Sigdef\ 
where $V_D$ and $Q V_D$ are defined in \vd\ and \qvd. 

Using the BRST-invariant prescription of \twoa\ for the two-point amplitude involving charged bosons, the first term 
\eqn\firstr{\int d^2 \tau\langle \bar c V^{(1)}(z_1) \int_\e d^2 z_2 U^{(2)}(z_2)  ~{\cal N}(y) |(\int \mu b)|^2 \rangle}
is proportional to $k^2$ and therefore vanishes when the states are onshell. Indeed, in $d=10$, the charged boson is a gauge field $A_{\bar J}$ which can only appear in
the gauge-invariant combination $F_{m \bar J} = \p_m A_{\bar J} - \p_{\bar J} A_m$.
After dimensional reduction, this implies that the two-point amplitude must be proportional to 
$\eta^{mn} g^{J \bar J} F_{m \bar J} F_{n J} = k^2 g^{J \bar J} A_J A_{\bar J}$.
However, the second term of \twoa,
\eqn\twor{\int d^2 \tau \langle \bar c ( - \d_{T T'} V_D + \dots) ~{\cal N}(y) |(\int \mu b)|^2 \rangle}
is nonvanishing, as we will argue in the next section.

For charged fermions in the same supersymmetric multiplet as the bosons, Lorentz invariance and charge conservation imply that the two-point amplitude must be proportional to
$\g_Q^{\a\b} k^Q$. But $\g_Q^{\a\b} k^Q$ vanishes when the states are onshell, 
so the
two-point amplitude for charged fermions is zero and there is a one-loop mass splitting if the
one-loop mass shift of the bosons is non-zero.

\subsec{One-loop expectation value}

As explained in \ref\mafra{ N.~Berkovits and C.~R.~Mafra,
``Some Superstring Amplitude Computations with the Non-Minimal Pure Spinor Formalism,''
JHEP {\bf 0611}, 079 (2006).
[hep-th/0607187].}, the one-loop open superstring amplitude in the
pure spinor formalism is computed by
\eqn\oneloop{ A = \int d \tau  \langle (\int b) {\cal N}(w) V(z_1) 
\prod_{r=2}^N \int dz_r U(z_r) \rangle }
where $b$ is the composite operator satisfying $\{Q, b\} = T$ and
\eqn\regul{{\cal N} = e^{-\lambda^\alpha\bar\lambda_\alpha -\theta^\a r_\alpha - w_\alpha\bar w^\alpha - s^\alpha d_\alpha}}
is the BRST-invariant regulator.  (In genus one, it is convenient to represent one vertex operator in unintegrated form and the others
in integrated form.)
Although one here wants to do the one-loop heterotic superstring amplitude
computation, the heterotic computation can be easily obtained from the
open superstring computation by adding the right-moving 
$(\bar b,\bar c)$ ghosts and multiplying all super-Yang-Mills
vertex operators by the right-moving current $\overline J^I= \xi^K T^I_{K\bar K} 
\xi^{\bar K}$. 

In a flat background, it is easy to show that 
$\langle V_D \rangle=0$ because
one cannot absorb all the fermionic zero modes of $\theta^\alpha$ and their conjugate
momenta $p_\alpha$. However,
in a Calabi-Yau background, some of these fermionic zero modes can come from
the background field-strengths $F_{J\bar K}$ which contribute
$\int d^2 z F_{J\bar K}^i (\theta\gamma^{J\bar K} p) \overline J^i$ to the worldsheet
action through their integrated vertex operator $\int d^2 z U^{J\bar K}(z)$.

Using the zero mode analysis of \mafra, it is easy to verify
that the contribution 
to $\langle V_D \rangle$ to lowest-order in $\alpha'$ comes if three factors of the $U^{J\bar K}$ vertex
operator contribute from the worldsheet action. So instead of computing
the one-point amplitude $\langle V_D \rangle$ in a Calabi-Yau background, one can
instead compute the four-point amplitude
\eqn\aflat{A =\int d\tau \langle (\int \bar b) (\int  b) {\cal N}(w) V_D(z_1) \prod_{r=2}^4 \int d^2 z_r U^{J\bar K}(z_r) \rangle}
in a flat background. As in the four gluon one-loop amplitude computation in
\mafra, the only contribution to \aflat\ comes if $V_D$ contributes 
three $\theta^\alpha$
zero modes, each $U^{J\bar K}$ contributes one $\theta^\alpha$ 
zero mode and one $p_\alpha$ zero mode, 
the $b$ ghost contributes one $r_\alpha$ zero mode and two $p_\alpha$
zero modes, and ${\cal N}$ contributes 10 $\theta^\alpha$ zero modes, 
10 $r_\alpha$ zero modes, 11 $s^\alpha$ zero modes, and 11 $p_\alpha$ zero modes.

To compute the index contractions of \aflat,
it is
useful to note that $V_D$ of \vd\ resembles the term of order $(\theta)^3$ in
the vertex operator of an on-shell gluon with constant field-strength
in the direction $F_{J\bar J}=g_{J\bar J}$.
In ten uncompactified dimensions, the massless string field at ghost-number one
has the form  
\eqn\massless{
V = \lambda^\alpha A_\alpha (x,\theta)}
where $V$ only depends on the worldsheet zero modes. $QV=0$ and $\delta V = Q \Omega$ implies that $A_\a(x,\t)$ is the on-shell super-Yang-Mills spinor gauge field which
can be gauge-fixed to the form
\eqn\expand{A_\alpha(x,\theta) =\half a_m (x) (\gamma^m\theta)_\alpha +{1\over 3}  \xi^\beta(x) 
(\gamma^m \theta)_\alpha (\gamma_m\theta)_\beta -{1\over{32}} F_{np}(x) 
(\gamma_m\theta)_\alpha (\theta \gamma^{mnp}\theta) + ...}
where $...$ includes components which are related to derivatives of the
on-shell gluon 
$a_m(x)$ and gluino
$\xi^\beta(x)$ and
\eqn\fs{F_{mn} = \partial_m a_n -\partial_n a_m .}
 From \expand, the vertex operator
of an on-shell gluon with constant field-strength $F_{mn}$ is 
\eqn\vgluon{V = F_{mn} [(\lambda \gamma^m \theta) x^n +{1\over 8} (\lambda\gamma_q\theta)
(\theta\gamma^{qmn} \theta) ],} 
so the term without explicit $x^m$ dependence is the $V_D$ vertex operator
at zero momentum.
The same is true for \vdrns\ in the RNS formalism where the
on-shell gluon vertex operator is
\eqn\vgluonrns{V = F_{mn} [(\gamma \psi^m + 
c \partial x^m) x^n + c\psi^m\psi^n].}

Since by zero mode analysis, the $x$-dependent term in \vgluon\ does not
contribute to the four gluon one-loop computation in \mafra, the computation of
\aflat\ is identical to the computation of the four gluon scattering amplitude
in \mafra\ if one associates $V_D$ with the vertex operator of a gluon
with constant field-strength $F_{J\bar J} = D g_{J\bar J}$.
The index contraction of the four-gluon computation was shown in \mafra\
to reproduce the expected $t_8 \Tr (F^4)$ term, 
so one simply has to replace the indices of one of the $F$'s with 
$D g_{J\bar J}$
and replace the other three $F$'s with the Calabi-Yau background
field strengths $F_{J\bar K}$. Performing this replacement, one finds
that the $t_8 \Tr(F^4)$ term gives the index contraction
$D \Tr_{SU(3)} (F \wedge F \wedge F) $ as expected. 
So to lowest-order in $\alpha'$, the one-loop expectation value is $\langle V_D \rangle =p$
where $p$ is defined in \pinst.

In the RNS formalism, one can perform a similar analysis by working in
light-cone gauge and twisting the eight light-cone $\psi^j$ variables to have
integer spin (as in light-cone Green-Schwarz) so that they become 
periodic variables
on the worldsheet. In this case, one needs to absorb eight $\psi^j$ zero modes,
so the RNS vertex operator of \vdrns\ can have a non-zero one-loop 
amplitude only if the Calabi-Yau background 
field-strengths contribute three factors
of $\int dz F_{J\bar K} \psi^J \psi^{\bar K}$. As in the pure spinor
computation, the $\langle V_D \rangle$ computation is related to the
four gluon computation since zero mode analysis implies
that one can ignore the $x$-dependent terms in \vgluonrns\
and replace the unintegrated gluon vertex operator in the
four gluon one-loop computation with \vdrns.

In \DIS, the result $\langle V_D \rangle =p$ was argued to be exact to all orders in $\a'$ by using the light-cone Green-Schwarz formalism to show that massive string states do not contribute to $\langle V_D \rangle$. In the covariant RNS formalism, a similar argument can be made after choosing a special gauge for the worldsheet supermoduli such that the covariant RNS computation reduces to the light-cone computation. It would be interesting to look for a similar argument using the pure spinor formalism, perhaps involving a special choice for the regulator ${\cal N}$.  The computation
that we have presented here is only valid to lowest order in $\alpha'$.

\subsec{Simpler example of mass splitting}

As discussed in \Wone, the one-loop two-point function of a dilatino and a gaugino also exhibits a supersymmetry-breaking mass shift.  In the RNS formalism,
this supersymmetry-breaking effect is more straightforward 
 than the
example treated above:
many of the usual subtleties of superstring perturbation theory do not arise,  because the relevant super moduli space has only one odd modulus.
As we will now show,  this example is also more straightforward in the pure spinor approach, in this
case because it  can be computed  without introducing any cutoffs.

The pure spinor prescription for the one-loop dilatino-gaugino  two-point amplitude is  
\eqn\thr{A^\a_\b =
 \int d^2 \tau \langle \bar c V^{(1)\a}(z_1) \int d^2 z_2 U^{(2)}_\b(z_2) ~{\cal N}(y) |(\int \mu b)|^2 \rangle}
where $V^{(1)\a} = \g_n^{\a\d} (\l\g_m\t)(\g^m\t)_\d \bar\p x^n$ is the unintegrated vertex
operator for the dilatino at zero momentum, and 
$U^{(2)}_\b = (p_\b +\half (\p x^m +{1\over {12}} (\t\g^m \p\t)) (\g_m \t)_\b  )   (g_{J\bar J} \xi^J \xi^{\bar J})$ is the integrated vertex operator for the gaugino at zero momentum.

After contracting the $\bar\p x^n$ in $V^{(1)\a}$ with the $\p x^m$ in $U^{(2)}_\b$ and integrating over $d^2 z_2$, one
obtains
\eqn\thra{A^\a_\b = 
\int d^2 \tau \langle \bar c  (\l\g_m\t)(\g^n\g^m\t)^\a (\g_n\t)_\b   (g_{J\bar J} \xi^J \xi^{\bar J})~{\cal N}(y) |(\int \mu b)|^2 \rangle,}
which is proportional to
\eqn\thrb{\langle \bar c V_D (\g^{0123})_\b^\a ~{\cal N}(y) |(\int \mu b)|^2 \rangle =
 (\g^{0123})_\b^\a~  p }
where $V_D$ is defined in \vd\ and $p$ is defined in \pinst.

\subsec{ SUSY goldstino}

Here we will explore what goes wrong with a direct attempt to prove space-time supersymmetry.
(This discussion will be a pure spinor analog of the rather intricate
RNS discussion given in section 4 of \Wone.)
The naive computation to show that the one-loop mass shift is the same for supersymmetric
partners involves looking at $A = (\gamma^{0123})^\b_\a A^\a_\b$ where $\a$ and $\b$
are spinor indices which are SU(3) singlets and 
\eqn\twob{A_\b^\a =
 \int d^2 \tau \langle  \bar c ~q_\b ( V^{(1)}(z_1) \int d^2 z_2 U^{(2)\a}(z_2)) ~{\cal N}(y) |(\int \mu b)|^2 \rangle}
where $\bar c V^{(1)}$ is the unintegrated vertex operator for the boson, $U^{(2)\a}$ is the
integrated vertex operator for the fermion, and  $q_\beta$ is the susy charge whose
contour is integrated around the two vertex operators. 
The right hand side of \twob\ is the sum of two contributions in which $q_\b$ is commuted with one vertex operator or the other.
The sum of these two contributions is the difference of boson and fermion mass shifts, so $A=0$ is the condition for unbroken supersymmetry of these
amplitudes.  To try to prove that $A=0$, 
we pull the contour of $q_\b$ off of the vertex operators.  
The only contribution comes from the commutator of $q_\b$ with the regulator ${\cal N}$ of \call, which is  $[q_\b, {\cal N}] =- \rho~ r_\b {\cal N} =   Q(\rho~ \lb_\b {\cal N}).$
So if $ ( V^{(1)}(z_1) \int d^2 z_2 U^{(2)\a}(z_2)) $ were BRST-closed,
the BRST operator $Q$ could be pulled off of $(\lb_\b {\cal N})$ and $A_\b^\a$ would vanish. 

However, as discussed in subsection 3.1, 
$ ( V^{(1)}(z_1) \int d^2 z_2 U^{(2)\a}(z_2)) $ is not BRST-closed and satisfies
$Q (V^{(1)}(z_1) \int d^2 z_2 U^{(2)\a}(z_2)) = \Omega^\a(z_1)$ where
\eqn\defOma{\Omega^\a(z_1) = - V^{(1)}(z_1) \int_C d\bar z_2 V^{(2)\a}(z_2).}
If $\Omega^\a$
were equal to $Q\Sigma^\a$ for some $\Sigma^\a$, one could define a regularized version of 
\twob\ as in \twoa\ to be
\eqn\twoc{A_\b^\a =
 \int d^2 \tau \langle \bar c ~  q_\b ( V^{(1)}(z_1) \int_\e d^2 z_2 U^{(2)\a}(z_2) - \Sigma^\a(z_1)) ~{\cal N}(y) |(\int \mu b)|^2 \rangle.} 
Since 
$Q( V^{(1)}(z_1) \int_\e d^2 z_2 U^{(2)\a}(z_2) - \Sigma^\a(z_1))=0$, the argument of the previous
paragraph can be used to show that \twoc\ vanishes. 

But unlike the case in subsection 3.1, $\Omega^\a$ is a spacetime spinor and there does exist a spacetime spinor vertex operator in the 
BRST cohomology of ghost-number $(2,0)$, namely 
the dual vertex operator to the gluino,
\eqn\gluino{ V^{(3)\a} = (\l\g^m\t)(\l\g^n\t)(\g_{mn}\t)^\a  (g_{J\bar J}\xi^J \xi^{\bar J}) .}
This vertex operator $V^{(3)\a}$ is dual to the vertex operator of ghost-number $(1,0)$ for
the SUSY goldstino 
$V^{goldstino}_\a =  (\l\g^m\t)(\g_{m}\t)_\a  (g_{J\bar J}\xi^J \xi^{\bar J})$.

So if $\Omega^\a$ contains the vertex operator $V^{(3)\a}$, i.e. if the OPE of
$V^{(1)}$ and $V^{(2)\a}$  contains $V^{(3)\a}$ of \gluino,  $\Omega^\a$ is not BRST-trivial and one cannot write 
$\Omega^\a = Q\Sigma^\a$. Defining $\Omega^\a = Q\Sigma^\a + h V^{(3)\a}$
for some constant $h$, one finds that 
\eqn\twoe{A_\b^\a =
 \int d^2 \tau \langle \bar c ~  ( V^{(1)}(z_1) \int_\e d^2 z_2 U^{(2)\a}(z_2) - \Sigma^\a(z_1)) ~q_\b ({\cal N}(y)) |(\int \mu b)|^2 \rangle} 
$$
= \int d^2 \tau \langle \bar c ~  ( V^{(1)}(z_1) \int_\e d^2 z_2 U^{(2)\a}(z_2) - \Sigma^\a(z_1)) ~Q (\rho~\lb_\b {\cal N}) |(\int \mu b)|^2 \rangle$$ 
$$
= \int d^2 \tau \langle \bar c ~  Q ( V^{(1)}(z_1) \int_\e d^2 z_2 U^{(2)\a}(z_2) - \Sigma^\a(z_1)) ~(\rho~\lb_\b {\cal N}) |(\int \mu b)|^2 \rangle$$
$$
= h \int d^2 \tau \langle \bar c ~  V^{(3)\a}~ (\rho~\lb_\b {\cal N}) |(\int \mu b)|^2 \rangle.$$
Performing the functional integral over the pure spinors using the
regulator of \call, one obtains that $A^\a_\b$ is proportional to  $h (\g^{0123})^\a_\b$.

So the difference of the mass-shift computed by $A = (\g^{0123})_\a^\b A^\a_\b$ is proportional to $h$, which measures the coupling
of the bosonic and fermionic states described by $V^{(1)}$ and $V^{(2)\a}$ with the SUSY goldstino whose dual vertex operator $V^{(3)\a}$ is defined in \gluino.

\newsec{Two-Loop Cosmological Constant}

In the RNS framework, a natural procedure to evaluate the two-loop dilaton tadpole or cosmological constant
is to integrate over the odd moduli keeping fixed the super period matrix \DhPold.  This procedure actually gives
what one might call the bulk contribution to the cosmological constant.  In general, as explained in section 3 of \Wone,
one also requires a boundary contribution.\foot{This splitting between bulk and boundary contributions depends on the special
facts about super Riemann surfaces of genus 2 that were exploited in \DhPold.  Many of these facts generalize to genus 3 but not beyond.
For $g>3$,  there is a natural recipe to compute the genus $g$ cosmological constant, but it has no natural splitting as a sum of bulk and
boundary contributions.} In the case of a supersymmetric compactification to four dimensions, this boundary contribution
is a universal multiple of $\langle V_D\rangle^2$, where $\langle V_D\rangle$ is computed in genus 1.  It was conjectured in \Wone\ 
that in an arbitrary supersymmetric compactification to four (or more) dimensions, the bulk contribution to the genus 2 cosmological constant
vanishes and the full answer comes from the boundary contribution.    It is actually rather tricky to prove this in the RNS framework.
  The conjecture has been supported by explicit calculations in some orbifold compactifications
 to four dimensions with $N=1$ supersymmetry \DhP, and a general proof (based on an analysis of the super
 period matrix with Ramond punctures) will appear elsewhere.
 
 We will aim here to explore the analogous issues in the pure spinor formalism.  In this, we have to remember that in subsection 3.3,  we determined
 the one-loop expectation value $\langle V_D\rangle$ in the pure spinor formalism only to the lowest non-trivial order in $\alpha'$.   Similarly,
 at a certain point below, we will have to make an argument that is valid only to lowest non-trivial order in $\alpha'$.

In the pure spinor formalism, the naive formula for the two-loop cosmological constant is
\eqn\naivet{A =
 g_s^2\prod_{j=1}^3 \int d^2 \tau_j \langle {\cal N}(y) \prod_{j=1}^3 |(\int \mu_j b)|^2 \rangle}
 where ${\cal N} = e^{Q\L}$ is the regulator needed to perform the functional integral
 over the noncompact pure spinor zero modes and $\mu_j$ are the Beltrami differentials
 dual to the three complex Teichmuller parameters $\tau_j$ of the genus-two surface.
 
 As in the one-loop amplitude of the previous section, the possible presence of singularities implies that one needs to introduce cutoffs near the boundary of the moduli space of the genus-two surface.
 The genus-two surface can degenerate in two different ways: either by one of the two handles becoming infinitely thin or by the surface separating into two genus-one surfaces connected
 by an infinitely thin tube. The first type of degeneration is harmless for the same reason
 as in the one-loop amplitude; the momentum flowing through the degenerating handle is an integration variable. However, the second type of degeneration needs to be carefully treated and will give a contribution when spacetime supersymmetry is spontaneously broken at one-loop.
 
 We parametrize the genus 2 moduli space with parameters $\tau_1,\tau_2,\tau_3$ such that the separating degeneration occurs for 
  $\rm{Im}\,\tau_3\to\infty$   and that in this limit, $\tau_1$ and $\tau_2$ are the moduli of the two components.  (This may be done in a standard way
  with the $\tau_i$ being the matrix elements of the period matrix.)
The regularized expression for the vacuum amplitude requires a cutoff $(\rm{Im}\,\tau_3)^{-1}>\e$ on the modular integral, 
and one needs to verify that  with this cutoff in place, 
the amplitude remains independent of the choice of $\L$ in the regulator ${\cal N}=e^{Q\L}$. Under $\L \to \L + \d\L$, the amplitude prescription of  eqn. \naivet\ transforms as
\eqn\regtwo{\d A =
 g_s^2\int d^2 \tau_1 \int d^2 \tau_2 \int_\e d^2\tau_3 \langle (Q\d\L) {\cal N}(y) \prod_{j=1}^3 |(\int \mu_j b)|^2 \rangle}
$$=
g_s^2 \int d^2 \tau_1 \int d^2 \tau_2 \int_\e d^2\tau_3 \langle (\d\L){\cal N}(y)
(\int\mu_3 T)(\int\bar\mu_3 \bar b) \prod_{j=1}^2 |(\int \mu_j b)|^2 \rangle $$
$$=
 g_s^2\int d^2 \tau_1 \int d^2 \tau_2 \int_\e d^2\tau_3  {\p\over{\p\tau_3}} \langle (\d\L) {\cal N}(y)
(\int\bar\mu_3 \bar b) \prod_{j=1}^2 |(\int \mu_j b)|^2 \rangle $$
$$ = g_s^2 \sum_I \quad \int d^2 \tau_1 \langle (\d\L_1){\cal N}_1(y_1)  |(\int\mu_1 b)|^2 \bar c V^*_I(z_1)\rangle
\quad \int d^2 \tau_2 \langle  \bar c V_I(z_2) 
{\cal N}_2(y_2) |(\int\mu_2 b)|^2 \rangle $$
where the genus-two surface has been factorized into two genus-one surfaces connected by a thin tube and $\sum_I$ is a sum over all possible states going through the tube. 

It will be assumed that for $\rm{Im}\,\tau_3\to\infty$, the original regulator ${\cal N}$ factorizes as  the product of two regulators 
${\cal N}_1{\cal N}_2$ where ${\cal N}_1 = e^{Q\Lambda_1}$ is inserted on the first genus-one surface at $y_1$, ${\cal N}_2=e^{Q\Lambda_2}$ is inserted on the second genus-one surface at $y_2$, and $\L_1$ has been shifted by $\d\L_1$. This assumption is necessary since
when the surface degenerates, the two holomorphic zero modes of the conformal weight
one variables $(w_\a, d_\a, s_\a)$ separate onto
the two different surfaces. So the regulator of \call\ only provides the correct zero mode dependence if it factorizes into two regulators on the different surfaces. This is analogous
to the necessity in the RNS formalism of distributing the picture-changing operators on the different
surfaces.
  
In eqn. \regtwo, the state
going into the thin tube from one side is represented by the vertex operator $\bar c V^*_I(z_1)$ of ghost-number $(2,1)$ and the state going out of the thin tube on the other side
is represented by the vertex operator
$\bar c V_I(z_2)$ of ghost-number $(1,1)$. The vertex operators $\bar c V^*_I$
and $\bar c V_J$ are defined to satisfy the dual relation
\eqn\dualvs{\langle \bar c V^*_I(z_1) \bar c V_J(z_2) \rangle = \d_{IJ}}
using the tree-level measure factor of \measure.

As discussed in subsection 2.2, the one-point one-loop amplitude is nonzero
if the vertex operator $V_I$ is equal to $V_D$ of \vd. Furthermore, one can always choose a basis of vertex operators such that there is a unique vertex operator with 
non-vanishing one-point one-loop amplitude. For example, if two vertex operators have non-vanishing one-point one-loop amplitudes, one can construct a linear combination of these vertex operators with vanishing one-point one-loop amplitude. Assuming that $V_D$ is
this unique vertex operator, one obtains that 
\eqn\sod{\int d^2 \tau_2 \langle {\cal N}_2(y_2) \bar c V_I(z_2) 
|(\int\mu_2 b)|^2 \rangle = p 
\d_{ID}}
where $p$ is defined in \pinst.
And since
$QV_D$ of \qvd\ satisfies $\langle V_D (Q V_D) \rangle =1$ using the measure
factor of \measure, $V_D^* = QV_D$ is consistent with
\dualvs. So $\d A$ of \regtwo\ is equal to
\eqn\regso{ \d A = g_s^2\int d^2 \tau_1 \langle (\d\L_1){\cal N}_1(y_1)  |(\int\mu_1 b)|^2 \bar c V^*_D(z_1) \rangle
\quad \int d^2 \tau_2 \langle \bar c V_D(z_2){\cal  N}_2(y_2)  |(\int\mu_2 b)|^2 \rangle }
$$= 
g_s^2 \int d^2 \tau_1 \langle (\d\L_1){\cal N}_1(y_1)  |(\int\mu_1 b)|^2 \bar c QV_D(z_1)\rangle
\quad \int d^2 \tau_2 \langle \bar c V_D(z_2){\cal N}_2(y_2)  |(\int\mu_2 b)|^2 \rangle $$
$$= - g_s^2
\int d^2 \tau_1 \langle Q(\d\L_1){\cal N}_1(y_1)  |(\int\mu_1 b)|^2 \bar c V_D(z_1) \rangle
\quad \int d^2 \tau_2 \langle \bar c V_D(z_2) {\cal N}_2(y_2)  |(\int\mu_2 b)|^2 \rangle.$$

One can therefore define a manifestly BRST-invariant two-loop cosmological
constant with the prescription
\eqn\regtt{ A = g_s^2
 \int d^2 \tau_1 \int d^2 \tau_2 \int_{1/\rm{Im}\,\tau_3>\e} d^2\tau_3 \langle {\cal N}_1(y_1){\cal N}_2(y_2) \prod_{j=1}^3 |(\int \mu_j b)|^2 \rangle}
$$+
 g_s^2\int d^2 \tau_1 \langle {\cal N}_1(y_1) \bar c V_D(z_1) |(\int\mu_1 b)|^2 \rangle
\quad \int d^2 \tau_2 \langle {\cal N}_2(y_2) \bar c V_D(z_2) |(\int\mu_2 b)|^2 \rangle$$
$$ =
g_s^2 \int d^2 \tau_1 \int d^2 \tau_2 \int_{1/\rm{Im}\,\tau_3>\e}  d^2\tau_3 \langle {\cal N}_1(y_1) {\cal N}_2(y_2) \prod_{j=1}^3 |(\int \mu_j b)|^2 \rangle \quad + g_s^2 p^2,$$
where we have used the lowest order result for the one-loop expectation value $\langle V_D\rangle$.

Finally, it will be argued by analyzing zero modes of $\t^\a$
that the first term on the right hand side of \regtt\ vanishes to lowest non-trivial order in $\alpha'$.
Under the $SO(3,1)\times SU(3) \times U(1)$ decomposition of $SO(9,1)$, there
are four $\t^\a$ zero modes which are $SU(3)$ singlets and cannot come from
the worldsheet action of \curvedhet. So for the first term to be non-vanishing, these four zero modes must all come from the $e^{-\t^\a r_\a}$ term in the regulator ${\cal N}$ of \regul. In fact, the argument will only require that at least one $\t^\a$ zero mode must come from ${\cal N}$.

Suppose that $\t^{+}$ comes from ${\cal N}_1 {\cal N}_2$ in \regtt\ where, under a $U(2)\times SU(3)\times U(1)$ subgroup of the Wick-rotated $SO(10)$, $\t^+$ denotes the singlet with 
total $U(1)$ charge $5\over 2$. It will be convenient to decompose the Wick-rotated $SO(10)$ spinors $(\l^\a, \t^\a)$ and $(\lb_\a, \tb_\a)$ in 
$U(5)$ notation as \Bnon
\eqn\decom{\l^\a = \l^+ [1, u_{ab}, -{1\over 8}\e^{abcde} u_{bc} u_{de}], \quad
\t^\a = [\t^+, \t_{ab}, \t^a],}
$$\lb_\a = \lb_+ [1, \overline u^{ab}, - {1\over 8}\e_{abcde} \overline u^{bc} \overline u^{de}], \quad
r_\a = [r_+, r^{ab}+ r_+ \overline u^{ab},  -{1\over 8} (r_+ \overline u^{bc}\overline u^{de} +2 r^{bc}\overline u^{de})],$$
where $a=1$ to 5 and $(u_{ab}, \overline u^{ab})$ parameterize the compact coset $SO(10)/U(5)$.

Consider the correlation function  
\eqn\fixed{ 
  \langle f(\l^+ \lb_+ +\t^+ r_+) {\cal N}_1(y_1){\cal N}_2(y_2) \prod_{j=1}^3 |(\int \mu_j b)|^2 \rangle}
 $$= 
 \langle [f(\l^+ \lb_+) +f'(\l^+ \lb_+)(\t^+ r_+)] {\cal N}_1(y_1){\cal N}_2(y_2) \prod_{j=1}^3 |(\int \mu_j b)|^2 \rangle$$
$$= 
 \langle [f(\l^+ \lb_+) -f'(\l^+ \lb_+)] {\cal N}_1(y_1){\cal N}_2(y_2) \prod_{j=1}^3 |(\int \mu_j b)|^2 \rangle$$
where $f$ is an arbitrary function satisfying $f(0)=0$ and we have used that $(1+\t^+ r_+) {\cal N} = 0$ inside the correlation function since $(1+\t^+ r_+){\cal N}$ is independent of $\t^+$. Since $\l^+ \lb_+ + \t^+ r_+ = Q(\t^+ \lb_+)$, \eqn\defql{f(\l^+ \lb_+ +\t^+ r_+)=Q\Lambda}
for some $\Lambda$. So following the same arguments as in \regtwo\ and \regso,
\eqn\fixedtwo{
 \int d^2 \tau_1 \int d^2 \tau_2 \int_{1/\rm{Im}\,\tau_3>\e} d^2\tau_3 
  \langle  [f(\l^+ \lb_+) -f'(\l^+ \lb_+)]   {\cal N}_1(y_1){\cal N}_2(y_2) \prod_{j=1}^3 |(\int \mu_j b)|^2 \rangle}
  $$
=   \int d^2 \tau_1 \int d^2 \tau_2 \int_{1/\rm{Im}\,\tau_3>\e} d^2\tau_3 
  \langle  f(\l^+ \lb_+ +\t^+ r_+)  {\cal N}_1(y_1){\cal N}_2(y_2) \prod_{j=1}^3 |(\int \mu_j b)|^2 \rangle
$$
  $$
=   \int d^2 \tau_1 \int d^2 \tau_2 \int_{1/\rm{Im}\,\tau_3>\e} d^2\tau_3 
  \langle  (Q\Lambda)  {\cal N}_1(y_1){\cal N}_2(y_2) \prod_{j=1}^3 |(\int \mu_j b)|^2 \rangle
$$
$$
=\int d^2 \tau_1 \langle  \Lambda {\cal N}_1(y_1)~ \bar c QV_D(z_1) |(\int\mu_1 b)|^2 \rangle
\quad \int d^2 \tau_2 \langle {\cal N}_2(y_2) \bar c V_D(z_2) |(\int\mu_2 b)|^2 \rangle$$
 $$ = -
  \int d^2 \tau_1 \langle  f(\l^+ \lb_+ +\t^+ r_+) {\cal N}_1(y_1)~ \bar c V_D(z_1) |(\int\mu_1 b)|^2 \rangle
\,\, \int d^2 \tau_2 \langle {\cal N}_2(y_2) \bar c V_D(z_2) |(\int\mu_2 b)|^2 \rangle.$$

 The factor
$$ \int d^2 \tau_1 \langle  f(\l^+ \lb_+ +\t^+ r_+) {\cal N}_1(y_1)~ \bar c V_D(z_1) |(\int\mu_1 b)|^2 \rangle$$
that appears
 in the
last line of \fixedtwo\ can be analyzed in a manner familiar from
 subsection 3.3. In subsection 3.3, it was shown to lowest-order in $\a'$ that the vertex operator $V_D(z_1)$ can
be replaced by the BRST-invariant vertex operator of \vgluon, since zero mode analysis
implies that the $x$-dependent term in \vgluon\ does not contribute to the $F^4$ term in the one-loop computation. But since 
$ f(\l^+ \lb_+ +\t^+ r_+)$ is BRST-trivial, replacing $V_D$ with a BRST-invariant operator implies that the first term in the last line of \fixedtwo\ vanishes. So to lowest non-trivial order in $\a'$, the last line of \fixedtwo\ vanishes.
 
So we have 
shown that 
\eqn\fixedthree{
 \int d^2 \tau_1 \int d^2 \tau_2 \int_{1/\rm{Im}\,\tau_3>\e} d^2\tau_3 
  \langle  [f(\l^+ \lb_+) -f'(\l^+ \lb_+)]  {\cal N}_1(y_1){\cal N}_2(y_2) \prod_{j=1}^3 |(\int \mu_j b)|^2 \rangle =0}
 for any function $f$ satisfying $f(0)=0$. This is only possible if
after performing the functional integral over all worldsheet variables except for
$\l^+$ and $\lb_+$, 
\eqn\fixedfour{
  \langle  {\cal N}_1(y_1){\cal N}_2(y_2) \prod_{j=1}^3 (\int \mu_j b) \rangle 
= c  e^{-\l_+ \lb^+}}
for some proportionality constant $c$.

The functional integral measure for the pure spinors of \decom\ is 
\eqn\functm{\int d\l^+ d\lb_+ \int d^{10} u_{ab} \int d^{10} \overline u^{ab}  ~(\l^+ \lb_+)^{10},}
 so \fixedfour\ implies that $ \prod_{j=1}^3 (\int \mu_j b)$ diverges like $(\l^+\lb_+)^{-10}$ when $(\l^+\lb_+) \to 0$. But the maximum divergence of the $b$ ghost in \Bnon\ is
$(\l^+\lb_+)^{-3}$, so it is not possible for 3 $b$ ghosts to cancel the $(\l^+\lb_+)^{10}$ measure factor. So $c=0$ in \fixedfour, and the first term on the right-hand side of \regtt\ vanishes.

So we have shown that after including the contact term required
for BRST-invariance, the two-loop cosmological constant is equal to $ g_s^2 p^2 $, at least to lowest non-trivial 
order in $\a'$. This result is expected to hold to all orders in $\a'$ and it would be desirable to find a proof using the pure spinor formalism.

\vskip 10pt
{\bf Acknowledgements:}
We would like to thank Nathan Seiberg and Cumrun Vafa for useful discussions.
NB would like to thank
CNPq grant 300256/94-9
and FAPESP grants 09/50639-2 and 11/11973-4 for partial financial support.
Research of EW was partly supported by NSF Grant PHY-1314311.
\listrefs
\end